\documentclass[twocolumn,showpacs,preprintnumbers,amsmath,amssymb]{revtex4}
\usepackage{graphicx}
\begin{document}
\title{New High-$T_c$ Half-Heusler Ferromagnets NiMnZ (Z = Si, P, Ge, As)}
\author{Van An~\textsc{DINH}}
\email{divan@cmp.sanken.osaka-u.ac.jp.} 
\author{Kazunori ~\textsc{Sato} and Hiroshi~\textsc{Katayama-Yoshida}}

\affiliation{%
The Institute of Scientific and Industrial Research, Osaka University, Mihogaoka 8-1, Ibaraki, Osaka 567-0047, Japan.
}
\begin{abstract}
 Based on the first principle calculation, we propose a new class of high-$T_c$ half-heusler ferromagnets NiMnZ (Z = Si, P, Ge, As). The structural and magnetic properties are investigated through the calculation of the electronic structure, phase stability, equilibrium lattice constant,  magnetic exchange interaction $J_{ij}$ and  Curie temperature $T_c$. It is found that all alloys show half-metallicity and ferromagnetism at temperatures much higher than room temperature in a wide range of lattice expansion (compression). At the equilibrium lattice constant,  $T_c$ of 715K, 840K, 875K and 1050K are predicted by Monte Carlo simulation for NiMnP, NiMnAs, NiMnGe and NiMnSi, respectively. Following these results, these alloys are strongly expected to be promising candidates for spintronic applications.
\end{abstract}
\maketitle
%
\section{Introduction}
 The prediction and development of new magnetic materials and half metallic ferromagnets are necessary for exploiting the great potentials of spintronics in the near future. From the view point of industrial application, since the electronic structure behaves like metals w.r.t. the electrons of one spin direction and like semiconductors w.r.t. the spins in the opposite direction (also called half-metallicity), leading to 100\%
  spin polarization at Fermi level $E_F$, the half-metallic ferromagnet with Curie temperature ($T_c$) higher than room temperature is considered as a key material for spintronic applications.
   
\begin{figure} 
\begin{center}\leavevmode
\includegraphics[width=0.45\linewidth]{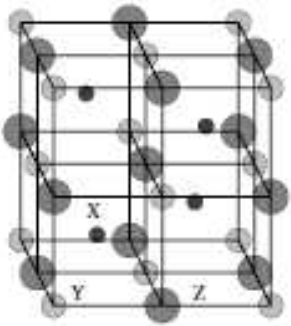}
\caption{$C1_b$ structure: the unit cell contains Y at $4a(0,0,0)$, X located in the octahedral coordinated pocket $4c(\frac{1}{4},\frac{1}{4},\frac{1}{4})$,  Z at the body center $4b(\frac{1}{2},\frac{1}{2},\frac{1}{2})$ and a vacant site at $4d(\frac{3}{4},\frac{3}{4},\frac{3}{4})$.\label{struct} }
\end{center}

\end{figure}

  The family of half-Heusler alloys NiMnZ (Z is an $sp-$valence element) that have $C1_{b}$ structure as illustrated in Fig.~\ref{struct}  are also greatly expected to be a ferromagnetic half-metal because the interaction between the $d$ orbitals of Ni and Mn (with suitable Z elements such as Sb) can form a gap between the bonding and anti-bonding states\cite{galanakis}. The half-metallicity is first predicted in the half-Heusler alloy NiMnSb by de Groot and collaborators\cite{groot} in 1983. Experimentally, besides the fact that the strong spin polarization has been verified in some compounds\cite{han1, han2,soulen}, the ferromagnetism in the half-Heusler NiMnZ at temperatures ranging between 500K and 730K has been known for Z= Co, Pd, Pt and Sb\cite{webster}. Theoretically, the attempts to replace Ni (or Mn) of NiMnSb by other metals such as Fe, Co, etc. give a half-metallic picture in the band structure of some alloys\cite{galanakis,nanda} such as FeMnSb, CoMnSb and FeCrSb, etc.
   
   In this paper, based on the first principle calculation, we propose a new class of high-$T_c$ ferromagnets NiMnZ (Z = Si, P, Ge, As) by placing the atoms lighter than Sn such as Si, P, Ge and As into Z sites of NiMnZ alloy. 
    For this aim, we have carried out the calculation as follows. First, we perform the total energy calculation to estimate the equilibrium lattice constants (ELC) and configuration stability of C$1_b$ structure by employing the ultrasoft pseudo-potential (UPP) method as the implement in STATE-Senri code (thanks to Prof Y. Morikawa, Osaka Univ.). 
     To evaluate ELC, we employ the generalized gradient approximation (GGA) within UPP as well as both of the muffin-tin (MTA) and atomic sphere approximations (ASA) within framework of LSDA (thanks to AKaiKKR package coded by Prof. H. Akai (Osaka Univ.))[Sec. 2]. Using the obtained ELCs, we next calculate the density of states (DOS) [Sec. 3], magnetic exchange interaction $J_{ij}$ between magnetic sites [Sec.4].  Finally, using $J_{ij}$ as the input data we calculate $T_c$ by employing three approaches: mean field approximation (MFA), random phase approximation (RPA) and Monte Carlo simulation (MC) [Sec. 5]. The effect of the lattice expansion/compression on the half-metallicity, magnetic exchange interaction and Curie temperature is also investigated. The structural and magnetic properties of the prototypical half-heusler alloy NiMnSb are also calculated for comparison. 
     
\section{Phase Stability and Equilibrium Lattice Constant}  

   We have performed the total energy calculation using GGA within the framework of UPP method to estimate the stability of three different configurations of NiMnZ: $\alpha$, $\beta$ and $\gamma$ phases. These phases are defined  in Table~\ref{tab0}. 
\begin{table}
\caption{Structural configurations of NiMnZ.}
\vskip 0.1cm
\label{tab0}
\begin{tabular}{@{\hspace{\tabcolsep}%
				\extracolsep{\fill}}llll} \hline
Phase & $4a-$site & $4b-$site& $4c$-site\\ 
 \hline
$\alpha$ &Mn &$sp-$element&Ni\\\hline
$\beta$&Mn&Ni&$sp-$element\\\hline
$\gamma$&Ni&$sp-$element&Mn\\\hline
\end{tabular}
\end{table}
The total energy differences between these phases show that the $\alpha$ phase is the most stable configuration and the $\beta$ phase is the least stable one. For example, total energy differences for NiMnSi are given as: $E(\beta)-E(\alpha)= 2.813$~eV and $E(\gamma)-E(\alpha)=1.11$~eV. Following this result, we next study ELC and magnetic properties of the alloys NiMnZ in the $\alpha$ phase.
   
 
\begin{table}
\caption{Equilibrium lattice constants calculated by employing the MTA ($a_{MT}$) and ASA ($a_{AS}$) within KKR-LSDA, and by GGA within UPP ($a_{UP}$). Experimental value  taken from Ref.~\ref{exp} is given in the parenthesis.}
\vskip 0.1cm
\label{tab1}
\begin{tabular}{@{\hspace{\tabcolsep}%
				\extracolsep{\fill}}llll} \hline
Alloy & $a_{MT}$ (a.u.) & $a_{AS}$ (au) & $a_{UP}$ (au)\\ 
 \hline
NiMnSi&10.4696 & 10.3291&10.2612\\\hline
NiMnP&10.5878&10.4376&10.2782\\\hline
NiMnGe&10.7447&10.5388&10.4448\\\hline
NiMnAs&10.8276&10.6526&10.5778\\\hline
NiMnSb&11.5085&11.3020 (11.2098)&11.0900\\\hline
\end{tabular}
\end{table}
 ELCs obtained by three approaches: GGA within UPP, MTA and ASA within KKR-LSDA are shown by Tab.~\ref{tab1}. It should be noted that GGA often gives ELC larger than LDA. However, the LDA within KKR approach in our calculation produces ELC larger than GGA combined in UPP method. For the same compound, $a_{MT}$ is the largest and $a_{UP}$ is the smallest. Generally, the biggest deviation of $a_{MT}$ (or $a_{UP}$) is approximately smaller than 3\%
  in comparison with $a_{AS}$. In particular, for the prototypical alloy NiMnSb, the experimental lattice constant ($a_{exp}=11.2098$ a.u.) is found to be in the range of $a_{UP}<a<a_{AS}$ and is close to the lattice constant obtained by ASA ($a_{AS}=11.3020$ a.u.) with a deviation smaller than 1\%.
    Also, one can expect that the real lattice constant of these proposed alloys might be arranged in the range from $a_{MT}$ to $a_{UP}$. 
    
\section{Density of States}
\begin{figure} 
\begin{center}
\includegraphics[width=\linewidth]{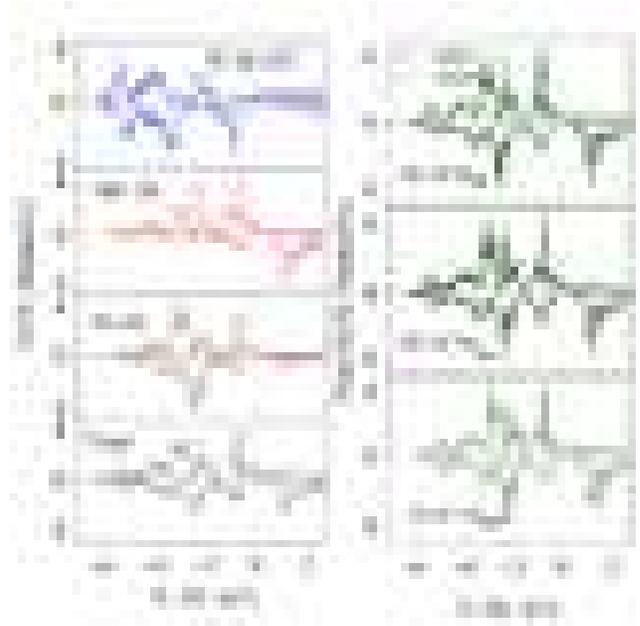} 
\caption{ DOS of NiMnSi. Left panel displays the partial DOS of $3p-$Si (multiplied by 5), $3d-$Ni and $3d-$Mn, and total DOS at $a=a_{AS}$. Right panel shows the total DOS within LSDA (KKR) and GGA (UPP) at three values of equilibrium lattice constant $a = a_{UP}$, $a_{AS}$ and $a_{MT}$.}
\label{DOSf} 
\end{center}
\end{figure}
\subsection{Half-Metallicity}
DOSs at ELCs are shown in Fig.~\ref{DOSf}. In this paper, only DOS of NiMnSi will be presented as a typical case. The  discussion of the half-metallicity in the remaining alloys will be presented elsewhere in more details. Moreover, one can also find a detailed explanation about the origin of the gap in various half-heusler alloys in Ref.~\ref{gana}. However, it should be noted that although the DOS of NiMnZ with Z= As, Ge and Si were also considered in Ref.~\ref{gana}, the authors simply applied the experimental lattice constant of NiMnSb, that should be much larger than the real lattice constants (10\%
 larger than $a_{AS}$ for NiMnSi, for example), for all of the alloys in question, resulting in the disappearance of the half-metallic character. On the contrary, using equilibrium lattice constants, we obtain the half-metallic character in the band structure of these alloys. The partial DOS (left panel) and total DOS (right panel) are calculated by LSDA and GGA at $a_{AS}$, $a_{MT}$ and $a_{UP}$. As seen from the left panel of Fig.~\ref{DOSf}, while most Ni-$3d$ states  concentrate at lower energies (in valence band) and are fully occupied, the $3d$ states of Mn are distributed to both lower and higher energies and are strongly polarized. The $3d-$majority spin states of Mn are located at lower energies and almost occupied. They hybridize with Ni-$3d$ states (and Si-$3p$ states) to form a band at $E_F$. The minority spin states of Mn shift to higher energies and are unoccupied, leading to a gap (where $E_F$ falls into) being formed in the minority band. Hence, the alloys are the half-metal. The calculated total magnetic moment $M_t$ of NiMnSi is exactly $3\mu_B$ in consistence with "rule of 18" $M_t=N_t-18$, where $N_t$ is the total number of valence electrons ($N_t$(NiMnSi)$ = 21$). Similarly, we have also obtained the half-metallic behavior and the total magnetic moment of $3\mu_B$ for NiMnGe, $4\mu_B$ for NiMnP, NiMnAs and NiMnSb at $a = a_{UP}$, $a_{AS}$ and $a_{MT}$ (except NiMnGe and NiMnSb at $a=a_{MT}$. See also Table~\ref{tab2}).
 \subsection{Influence of Lattice Expansion on DOS} 
  To study the effect of the lattice expansion (compression), we draw DOS of NiMnSi at different lattice constants in the right panel of Fig.~\ref{DOSf}. DOSs calculated at ELCs $a_{UP}, a_{AS}$ and $a_{MT}$ are plotted to illustrate the change of $E_F$ according to the increase of lattice constant $a$. The figure in the top panel corresponds to DOS at the smallest $a$, and the figure at the bottom illustrates DOS at the biggest $a$. In addition, in order to compare with LSDA, DOS calculated by GGA (dashed line) is also shown. The minority states of Mn-$3d$ electrons in GGA calculation shift toward the higher energies while the majority band is mostly kept unchanged, making the gap of the minority band wider than the LSDA. As the lattice constant $a$ increases(decreases), $E_F$ shifts toward the valence band (conduction band) and can fall into the valence (conduction) band, hence the half-metallicity might be destroyed if $a$ exceeds the threshold value as in the cases of NiMnSb and NiMnGe at $a_{MT}$. It is found that the half-metallicity of alloys NiMnSi, NiMnP and NiMnAs is preserved in the wider range of the expansion (compression) of the lattice cells.

\section{Magnetic Exchange Interaction}
  To investigate the ferromagnetism in NiMnZ alloys, we calculate the magnetic exchange interaction $J_{ij}$ at three values of ELC. Once $J_{ij}$ is obtained, the statistical methods such as MFA, RPA and MC are employed to calculate $T_c$. The exchange interaction $J_{ij}$ between two impurities at $ith$ and $jth$ sites, which are embedded in the ferromagnetic medium, is efficiently calculated by utilizing the magnetic force theorem. The frozen potential approximation \cite{oswald} is employed and Liechtenstein formula \cite{liech} is used for evaluation of $J_{ij}$ 
   \begin{equation}
   J_{ij}=\frac{1}{4\pi}\hbox{Im}\int^{E_F}{dE \hbox{Tr}\{(t_i^{\uparrow}-t_j^\downarrow)G^\uparrow_{ij}(E)(t^\uparrow_i-t^\downarrow_j)G^\downarrow_{ij}(E)\}},
   \end{equation}
 where $t_i^{\uparrow/\downarrow}$ are the site dependent scattering matrices for spin direction up ($\uparrow$) or down ($\downarrow$), and $G_{ij}^{\uparrow/\downarrow}$ are the KKR Green's functions connecting the $ith$ and $jth$ sites. The total energy change due to infinitesimal rotations of two magnetic moments at the $ith$ and $jth$ sites is calculated by using the magnetic force theorem, and the effective exchange interaction $J_{ij}$ is calculated via mapping the total energy change onto the classical Heisenberg model $H=-\Sigma_{i\ne j}{J_{ij}\mathbf{e}_i\mathbf{e}_j}$, where $\mathbf{e}_i$ denotes a unit vector parallel to the magnetic moment at the $ith$ site. 
 \subsection{Distance Dependence}
 \begin{figure}
\begin{center}
\includegraphics[width=0.90\linewidth, height=0.90\linewidth]{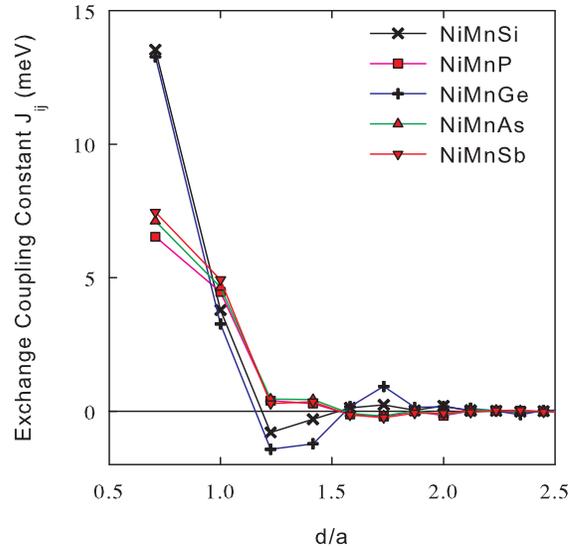}
\caption{Mn-Mn effective exchange coupling constant $J_{ij}$ vs. distance $d$ in units of lattice constant of NiMnZ (Z = Si, P, Ge, As, Sb). $J_{ij}$ is calculated at lattice constant $a=a_{AS}$. \label{jij}
}
\end{center}
\end{figure}
  Fig.~\ref{jij} displays $J_{ij}$ of NiMnZ at $a=a_{AS}$. In this figure, the influence of the $sp$-element on the magnetic exchange interaction is obvious. In alloys with $N_t=21$ (NiMnSi and NiMnGe), the exchange interaction is  strong and ferromagnetic for the $1st$ and the $2nd$ nearest neighbor pairs but anti-ferromagnetic for the $3rd$ and the $4th$ pairs. At farther distances, the interaction becomes weaker but ferromagnetic. In alloys having $N_t=22$ (NiMnP, NiMnAs and NiMnSb), the exchange interaction is ferromagnetic up to the $4th$ neighbor pairs but becomes weaker and anti-ferromagnetic at farther distances. In half-heusler alloys NiMnZ, $J_{02}$ of alloys having a bigger $N_t$ is stronger whereas $J_{01}$ is weaker.  A similar dependence can also be seen in CMR manganites\cite{soloviev1}.
 
  \subsection{Configuration Dependence}      
\begin{figure}
\begin{center}
\includegraphics[width=\linewidth, height=\linewidth]{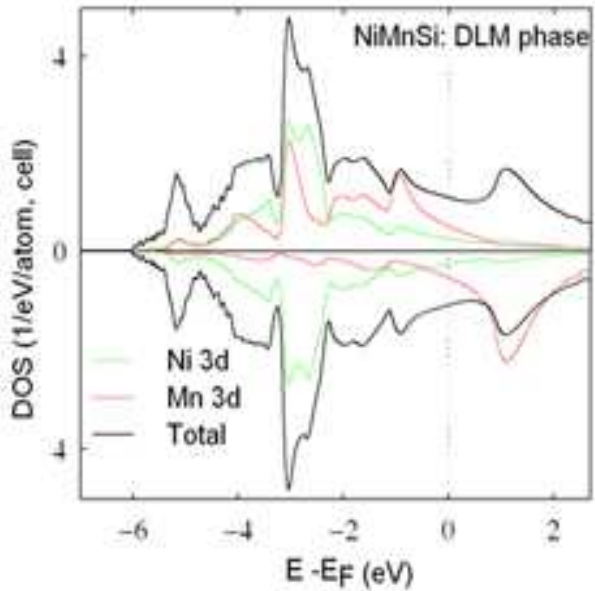}
\caption{DOS of DLM phase in NiMnSi.\label{DOSa} }
\end{center} 
\end{figure}

\begin{figure}
\begin{center}
\includegraphics[width=\linewidth, height=\linewidth]{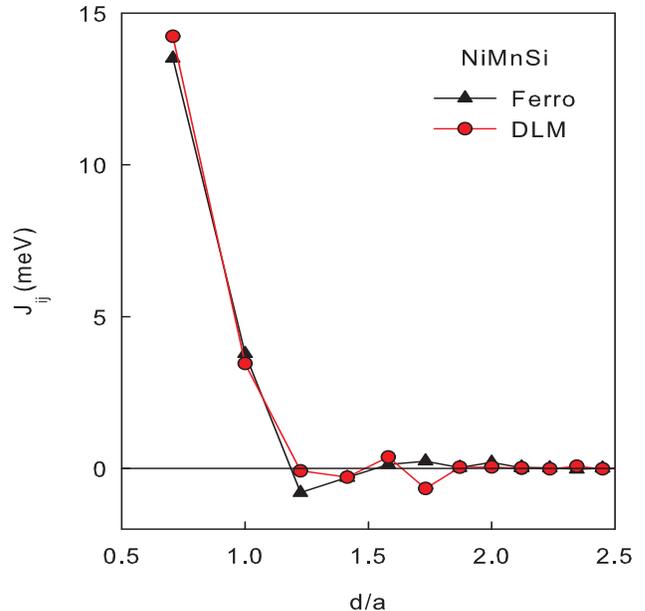}
\caption{Mn-Mn effective exchange coupling constant $J_{ij}$ vs. distance $d$ in units of lattice constant of NiMnSi. $J_{ij}$ is calculated at lattice constant $a=a_{AS}$ for the ferromagnetic (triangle) and disordered local moment (solid circle) states.\label{jijc}
}
\end{center} 
\end{figure}

The mapping on the Heisenberg model can produce different sets of parameters in the calculation of $J_{ij}$ of different magnetic configurations (say, ferromagnetic and antiferromagnetic phases) for CMR manganites\cite{soloviev2}. Thus, the evaluation of $T_c$ based on $J_{ij}$ sets can produce the different results for the same alloy. To consider the accuracy of this approximation in the calculation for our proposed half-heusler alloys, we now compare the electronic structure and magnetic exchange interaction of two phases: ferromagnetic and disordered local moment (DLM) phases. The DLM phase is simulated by the configuration in which half of the local magnetic moments point to the up direction and the others point to the down direction. DOS and $J_{ij}$ of DLM phase of NiMnSi are calculated in KKR-CPA-LSDA manner.  As seen from Fig.~\ref{DOSf} and Fig.~\ref{DOSa}, DOS of DLM phase is quite similar to the ferromagnetic phase with the majority spin states of Mn concentrating at lower energies and minority spins of Mn mostly distributing in the higher energy region. Consequently, DLM phase exhibits a behavior of $J_{ij}$ similar to the ferromagnetic phase with a small change in strength (Fig.~\ref{jijc}). $J_{ij}$ of DLM phase is stronger than that of ferromagnetic phase for the odd ($1st$, $3rd$ and $5th$) nearest neighbor pairs and weaker for the even ($2nd$, $4th$ and $6th$) nearest neighbor pairs. This differences is ignorable for the $1st$ and $2nd$ nearest neighbor pairs ($[J_{ij}^{\text{FM}}-J_{ij}^{\text{DLM}}]/J_{ij}^{\text{FM}}$ is very small). Note that the magnetic exchange coupling interaction in NiMnZ is short-ranged and very strong at very nearest distances and the dominant contributions into $T_c$ come from the $1st$ and $2nd$ nearest neighbors; therefore, this change of $J_{ij}$ does not affect $T_c$ so much. 
 \subsection{Effect of Lattice Expansion}    
\begin{figure}  
\begin{center}
\includegraphics[width=\linewidth]{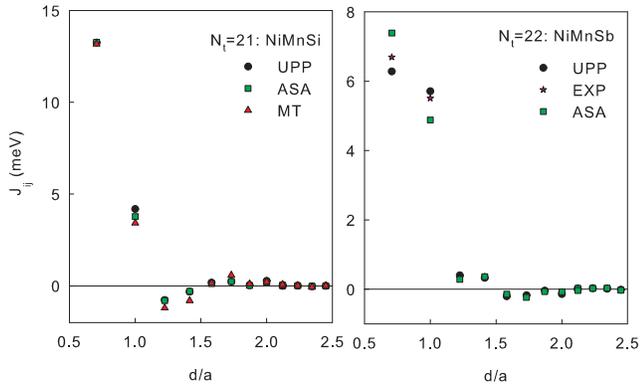}
\caption{$J_{ij}$ vs. lattice expansion of half-heusler alloys with $N_t=21$ (left figure) and $N_t=22$ (right figure). The triangles, squares, stars and circles denote $J_{ij}$ calculated at $a_{UP}$, $a_{AS}$, $a_{exp}$ and $a_{MT}$, respectively.\label{Jija}
}
\end{center}
\end{figure}

     Now let us consider the effect of lattice expansion on the exchange interaction $J_{ij}$. As shown in Fig.~\ref{DOSf}, the expansion lattice causes the shift of $E_F$ toward the valence band, hence it affects the exchange interaction between atoms on lattice sites. Fig.~\ref{Jija} illustrates the influence of the expansion lattice constant on the exchange interaction between Mn atoms of the fcc lattice. NiMnSi and NiMnSb are chosen here as the typical alloys corresponding to $N_t=21$ and $N_t=22$, respectively. $J_{ij}$ of alloys with $N_t=21$ (NiMnSi) and $N_t=22$ (NiMnSb) are calculated at some values of the lattice constant $a$. For all cases in question, $J_{0j}$ ($1<j<5$) decreases with increasing $a$. The change of $J_{0j}$ with higher $j$ is very small and does not affect $T_c$ so much. However, while $J_{01}$ of alloys with $N_t=21$ decreases and its change is ignorable, $J_{01}$ of alloys with $N_t=22$ remarkably increases with $a$. This leads to the different behavior of $T_c$ when $a$ varies: $T_c$ of alloys having 21 valence electrons decreases whereas that of alloys with 22 valence electrons increases with a lattice expansion. Such a variation of $T_c$ is shown explicitly in Table~\ref{tab2}.
  
\section{Curie Temperature}   
\begin{table}
\caption{$T_c$ calculated by MFA ($T_c^{MFA}$), RPA ($T_c^{RPA}$) and MC ($T_c^{MC}$) at $a = a_{MT}, a_{AS}$, $a_{UP}$ and experimental lattice constant $a_{exp}$. The experimental value\cite{webster} of $T_c$ is given in the parenthesis. 'HM' means half-metal. 
}
\vskip 0.1cm
\label{tab2}
\begin{center}
\begin{tabular}{@{\hspace{\tabcolsep}%
				\extracolsep{\fill}}lclrrrl} \hline
Alloy &$N_t$&$a$& $T_c^{MFA}$& $T_c^{RPA}$&$T_c^{MC}$&Physical\\ 
&&&(K)&(K)&(K)& nature\\
 \hline
 &&$a_{UP}$&1327 & 1027&1076& HM\\
NiMnSi&21&$a_{AS}$&1318& 1007&1050&HM\\
  &&$a_{MT}$&1198&892&968& HM\\\hline
  &&$a_{UP}$&1178&876&942& HM\\
NiMnGe&21&$a_{AS}$&1149&834&875&*HM\\
  &&$a_{MT}$&711& 561& & Metal\\\hline
  &&$a_{UP}$&746&606&625& HM\\
NiMnP&22&$a_{AS}$&863&699&715&HM\\
  &&$a_{MT}$&885&723&762& HM\\\hline
  &&$a_{UP}$&944 & 786&816& HM\\
NiMnAs&22&$a_{AS}$&973&807&840&HM\\
  &&$a_{MT}$&962& 802&867& HM\\\hline
  &&$a_{UP}$&877& 705&736& HM\\
NiMnSb&22&$a_{exp}$&898&716&745&HM\\
      &&&&&(730)&(HM)\\
  &&$a_{AS}$&928& 728&748& HM\\ 
  &&$a_{MT}$&839& 665& & Metal\\\hline
\end{tabular}
\end{center}
{\small *HM:\it Although the LSDA calculation shows the half-metallicity, the results obtained by GGA present a metallic behavior in the bands of both spin channels.}
\end{table}
 After calculating $J_{ij}$, the calculation of $T_c$ is carried out by performing three statistical approaches:  the mean field approximation, the random phase approximation and the Monte Carlo simulation. $T_c$ is defined as
 \begin{equation}
 k_BT_c^{MFA}=\frac{2}{3}\sum_{j}{J_{0j}} 
 \end{equation}
  in MFA, where $k_B$ is Boltzmann constant, and by\cite{hilbert,bouz}  
  \begin{equation}
  (k_BT_c^{RPA})^{-1}=\frac{3}{2N}\sum_{\bf q}{\left[J(0)-J({\bf q})\right]^{-1}},
  \end{equation}
   in RPA. Here, $J({\bf q})$ is the Fourier transform of the exchange parameter $J_{ij}$. 
   
   To perform MC simulation, the Metropolis algorithm\cite{binder} is applied to calculate the thermal average of the magnetization $M$ and its powers. Then, the Binder's cumulant crossing method\cite{binder} is employed and the fourth order cumulant $U_4$ defined by 
 \begin{equation}
  U_4=\displaystyle\frac{1}{2}\left\{ 5-3{\frac{<M^4>}{<M^2>^2}}\right\}
 \end{equation}
    is calculated as a function of temperature for different cell sizes ($14\times 14\times 14, 16\times 16\times 16$, and $18\times 18\times 18$ conventional fcc cells) to find the universal fixed-point at $T_c$. 
 
     In Tab.~\ref{tab2}, we compare the Curie temperature $T_c$ obtained by the Monte Carlo simulations with the results of MFA and RPA. It should be emphasized that the $T_c^{MC}$ of NiMnSb calculated at ELCs as well as at the experimental value is in very good agreement with the experimental $T_c=730$K\cite{webster}. It can also be seen that MFA overestimates $T_c$ even at 100\%
  magnetic atoms, whereas RPA underestimates $T_c$. Thus, as usual, the MC simulation, which can be considered as the most accurate method, gives the $T_c$ between values obtained by RPA and MFA at the same lattice constant. Also, the experimental $T_c$ of NiMnSb lies between the values calculated by RPA and MC and is close to $T_c^{MC}$. Hence, we will  hereafter refer to the $T_c^{MC}$ only. As expected, $T_c^{MC}$ of all alloys is much higher than room temperature and ranges from 715K to 1050K (at $a_{AS}$). The lowest $T_c$ corresponds to NiMnP with $T_c=715$K. NiMnSi has the highest $T_c$ ($1050$K at $a_{AS}$) and the next highest ones are NiMnGe (875K), NiMnAs (840K) and NiMnSb (730K at $a_{exp}$). Due to the $sp$-element, the variation of $T_c$ w.r.t. the expansion of lattice cells is different according to $N_t$. Alloys with $N_t=21$ have $T_c$ decreasing with $a$ (in the range where the half-metallicity is preserved), whereas alloys with $N_t=22$ have $T_c$ increasing. Accordingly, the half-metallicity in NiMnGe is considerably sensitive to the increase of the lattice constant due to $E_F$ being very close to the valence band edge. In contrast to LDA calculation which gets the half-metallicity at  $a=a_{AS}$, GGA calculation shows that the half-metallicity of NiMnGe can be destroyed if $a\ge a_{AS}$. However, similar to NiMnSb the real lattice constant of NiMnGe is expected to be smaller than $a_{AS}$ and then the half-metallicity might be preserved. In short, the half-metallicity in the alloys proposed (except NiMnGe) can be preserved in a wide range of lattice expansion.  
 \section{Conclusion}
In summary, we have investigated the structural and magnetic properties of the half-heusler alloys NiMnZ (Z = Si, P, Ge, As, Sb). The configuration stability is investigated. The equilibrium lattice constants are predicted by three approximations within the ultrasoft pseudo-potential method and KKR-LSDA as well. Using the equilibrium lattice constants, the magnetic exchange interaction is calculated and then the calculation of $T_c$ is performed by employing MFA, RPA and Monte Carlo simulation. The role of $sp$-elements as well as the influence of the lattice expansion (compression) on the half-metallicity and $T_c$ are also investigated. With the results obtained, we propose new half-metallic half-heusler alloys NiMnZ (Z = Si, P, Ge, As) which have $T_c$ much higher than room temperature in a wide range of the lattice expansion (compression). In particular, the present study shows that NiMnSi has very high $T_c$ and hence is one of the most promising materials for spintronic application and we expect that it can be synthesized soon. In addition, the high temperature half-metallicity and ferromagnetism can also be found in some Ni-based half-heusler alloys with Mn being replaced by Cr\cite{an}.

 \begin{acknowledgments}
  This research was partially supported by a Grant-in-Aid for Scientific Research in Priority Areas "Quantum Simulators and Quantum Design" and "Semiconductor Nanospintronics," a Grand-in-Aid for Scientific Research for young researchers, JST- CREST, NEDO-nanotech, the 21st Century COE, and the JSPS core-to-core program "Computational Nano-materials Design." We are grateful to Prof. H. Akai and Prof Y. Morikawa (Osaka Univ.) for providing us with the first principle calculation packages and to Prof. A. Yanase (Osaka Univ.) for helpful discussions.
\end{acknowledgments}

\end{document}